\documentclass[1p,number]{elsarticle}
\usepackage{graphicx} 
\usepackage{amssymb} \usepackage{amsmath}
\usepackage{enumerate}
\usepackage{rotating}
\usepackage{afterpage}
\usepackage{soul,color}
\usepackage{algorithm}
\usepackage[noEnd=false]{algpseudocodex}
\algrenewcommand\algorithmicrequire{\textbf{Input: }}
\algrenewcommand\algorithmicoutput{\textbf{Output: }}

\begin{document}
\title{Minimal cover of high-dimensional chaotic attractors by embedded recurrent patterns}

\author[1]{Daniel L. Crane}
\author[1]{Ruslan L. Davidchack\corref{cor1}}\ead{r.davidchack@leicester.ac.uk}
\author[1]{Alexander N. Gorban}\ead{a.n.gorban@leicester.ac.uk}
\cortext[cor1]{Corresponding author}
\affiliation[1]{organization={School of Computing and Mathematical Sciences, University of Leicester},
                city={Leicester}, postcode={LE1 7RH}, country={UK}}
                
\date{\today}

\begin{abstract}
We propose a general method for constructing a minimal cover of high-dimensional chaotic attractors by embedded unstable recurrent patterns. By minimal cover we mean a subset of available patterns such that the approximation of chaotic dynamics by a minimal cover with a predefined proximity threshold is as good as the approximation by the full available set. The proximity measure, based on the concept of a directed Hausdorff distance, can be chosen with considerable freedom and adapted to the properties of a given chaotic system. In the context of a spatiotemporally chaotic attractor of the Kuramoto--Sivashinsky system on a periodic domain, we demonstrate that the minimal cover can be faithfully constructed even when the proximity measure is defined within a subspace of dimension much smaller than the dimension of space containing the attractor. We discuss how the minimal cover can be used to provide a reduced description of the attractor structure and the dynamics on it.\end{abstract}


\maketitle

\noindent {\bf Highlights:}
\begin{itemize}
\item Algorithm for constructing minimal cover of chaotic attractor by recurrent patterns.
\item Minimal cover of the Kuramoto-Sivashinsky chaotic attractor is constructed.
\item Shadowing of chaotic trajectory by patterns within the minimal cover is demonstrated.
\item Markov-type model is constructed based on patterns within the minimal cover.
\end{itemize}


\section{Introduction}\label{sec:intro}
Unstable periodic orbits form the skeleton of chaotic attractors, with shorter orbits giving the overall structure, and longer orbits refining this skeleton in smaller neighbourhoods.  This property was first brought to light when Poincar\'{e} conjectured that any motion of a dynamical system can be approximated by means of those of periodic type~\cite{birkhoff1927periodic}. More rigorously, according to the Pugh closing lemma and Kupka--Smale theory, for a $C^1$-generic dynamical system, the set of hyperbolic periodic orbits is dense in the non-wandering set (see, e.g.,~\cite{hasselblatt2002handbook}). These properties of dynamical systems opened up a previously inaccessible avenue for breaching the frontier of chaotic dynamics using a periodic orbit-centric approach. In low-dimensional systems, this approach has seen considerable success, culminating in the development of the periodic orbit theory~\cite{ChaosBook}.  The approach can also work well for low-dimensional attractors of high-dimensional systems~\cite{procaccia1988complex,christiansen97nlin,lan2008unstable}; however when the dimensionality of the attractor is large (several positive Lyapunov exponents), even a qualitative description of the structure of the attractor becomes challenging.

In recent years a considerable amount of progress has been made in locating periodic orbits and other types of recurrent patterns in high-dimensional chaotic systems. (Following an example of Ref.~\cite{christiansen97nlin}, we use the term 'recurrent patterns' as a collective name for all types of special solutions embedded in the chaotic dynamics of high-dimensional systems, including equilibria, travelling waves, periodic and relative periodic orbits, as well as other identifiable solutions that can be used to describe the structure of chaotic dynamics. In turbulence research, such solutions are also referred to as 'exact coherent structures'~\cite{Waleffe01} or states~\cite{graham2021arfm}. Another commonly used term is 'invariant solutions' or 'states'~\cite{kawahara2012arfm}.)  Examples include the work of L\'{o}pez {\em et al.}~\cite{lopez2005relative}, who presented a method for finding relative periodic solutions for differential equations with continuous symmetries, and used this method to find relative periodic solutions of the Complex Ginzburg--Landau equation.  Hof {\em et al.}~\cite{hof2004experimental} found experimental evidence of the existence of travelling waves in turbulent pipe flow, in agreement with the numerical studies of Faisst and Eckhardt~\cite{faisst2003traveling} and Wedin and Kerswell~\cite{wedin2004exact}. Zoldi and Greenside~\cite{zoldi1998spatially} used a damped-Newton method to find unstable periodic orbits in the Kuramoto--Sivashinsky (KS) equation.  Cvitanovi\'{c} {\em et al.}~\cite{cvitanovic2010state} used multiple shooting and the Levenberg--Marquardt algorithm to locate over 60\,000 unstable recurrent patterns in the KS equation on a periodic domain. Due to the presence of discrete and continuous symmetries, these patterns include not only the traditional equilibria and periodic orbits, but also travelling waves and relative periodic orbits~\cite{lopez2005relative,cvitanovic2010state} (also called `modulated travelling waves' in~\cite{kevrekidis1990back}).  Recently, the toolbox for detecting exact coherent structures in turbulent flows has been expanded by the use of machine-learning methods~\cite{page2023arxiv}.

Now that large numbers of periodic orbits and other recurrent patterns can be located in high-dimensional chaotic systems, the question is: how can we use this information to describe the structure of the attractor and dynamics on it?  There is a growing body of research that attempts to describe high-dimensional chaotic dynamics in terms of embedded unstable periodic orbits or other invariant structures.  For weakly turbulent flows, examples are the works of Kawahara {\em et al.}~\cite{kawahara2012arfm}, Chandler and Kerswell~\cite{chandler2013jfm}, Budanur {\em et al.}~\cite{budanur2017jfm}, Graham and Floryan~\cite{graham2021arfm}.  Among more abstract models related to spatiotemporal chaos, of note is the work of Maiocchi {\em et al.}~\cite{maiocchi2024phd} on the Lorenz'96 model.  More generally, the knowledge of recurrent structures can be used for developing model-reduction and coarse-graining approaches to representing complex high-dimensional dynamical systems, e.g.,~via Markov chains~\cite{yalniz2021prl}, symbolic dynamics~\cite{yalniz2020cha}, principal manifolds~\cite{gorban2010ijns}, multiscale modelling~\cite{gorban2006book}.

When a large number of recurrent patterns is detected in a given chaotic system, it is clear that there is a lot of redundancy in the data: recurrent patterns which are `close' to one another in the phase space contain similar information about the structure and dynamics of the attractor in their neighbourhood.  This brings up a question of how to select a representative small subset of recurrent patterns that give essentially the same information about the attractor as the full available set.  In this work we adopt the following pragmatic approach (first outlined by the authors in~\cite{crane2016arxiv}): From the set of all available recurrent patterns, we select a subset which maximally covers the chaotic attractor with minimal redundancy. We call such a subset a \textit{minimal cover} of the attractor. The elements of this subset can then be used to describe the overall structure of the attractor.  In order to perform the selection, we need to define the notion of `closeness' (or proximity) of recurrent patterns to one another and to points on the attractor.  Generally, the specific choice of metric is not important, since the metric of the embedding space changes under topologically conjugate transformations, while remaining equivalent in the sense that, if $x$ and $y$ are two distinct points in the space with metric $d_1$ and $x' = \phi(x)$ is some conjugate transformation, then $0 < C_1 < d_2(x',y') / d_1(x,y) < C_2 < \infty$, where $d_2$ is the metric in the transformed space.  

The key feature that distinguishes `close' points from `distant' points in phase space is that trajectories originating from `close' points will remain close for some time (albeit exponentially diverging in the unstable directions).  This is usually true even if we view the dynamics in projection on a low-dimensional subspace.  Therefore, it is preferable to measure distance not between individual points in space, but between segments of trajectories.  In this context, as we demonstrate below, the notion of the Hausdorff distance~\cite{hausdorff1927mengenlehre} between sets (in this case between segments of trajectories) is useful.

\section{Minimal cover construction algorithm}\label{sec:algorithm} 
%
Let $\mathrm{P}=\left\{p_i\right\}_{i=1}^{N_\mathrm{P}}$ be a set of recurrent patterns listed in order of increasing complexity (e.g.,~equilibria followed by periodic orbits of increasing period) and let W $\subseteq$ P be the minimal cover of P we want to construct.  The minimal cover construction algorithm is presented in Algorithm 1, where $d(p_i,\mathrm{W})$ denotes the \textit{distance} from $p_i$ to W, which will be defined in Section~\ref{sec:hausdorff}.  

%
\begin{algorithm}\label{alg:cover1}
\caption{Minimal cover construction algorithm}
\begin{algorithmic}
\Require $\mathrm{P}=\left\{p_i\right\}_{i=1}^{N_\mathrm{P}}$; ~$\varepsilon > 0$
\State W $\gets \{p_1\}$
\For{$i = 2, \ldots, N_\mathrm{P}$ }
  \If{$d(p_i,\mathrm{W})>\varepsilon$ }
    \State W $\gets \mathrm{W} \cup \{p_i\}$
  \EndIf
\EndFor

\hspace*{-1em}\Output W
\end{algorithmic}
\end{algorithm}

Intuitively speaking, we only include a new recurrent pattern into the set W if it explores parts of the attractor that are not yet covered by the current W.

\begin{figure}[tb]
\begin{center}
\includegraphics[width=0.7\textwidth]{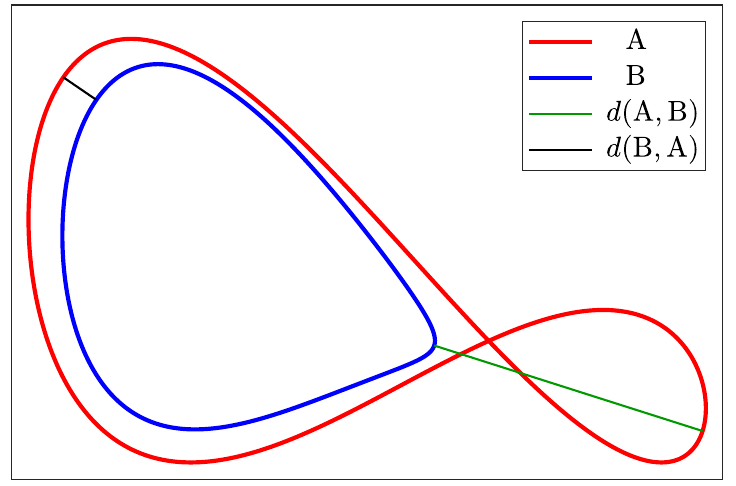}
\end{center}
\caption{\label{fig:dHdAB} Illustration of the directed Hausdorff distances between sets A and B in a two-dimensional plane with the Euclidean distance between points.}
\end{figure}

\subsection{Directed Hausdorff Distance}\label{sec:hausdorff}
Recurrent patterns $p_i(t), t \in [0, T^{p_i}]$, are sets of points in the dynamical system phase space parametrised by the time variable $t$.  In other words, they are segments of the dynamical system trajectories of duration $T^{p_i}$.  If $p_i$ is a periodic orbit, then $T^{p_i}$ is its period; if it is an equilibrium, then we take $T^{p_i} = 0$.  During the minimal cover construction process presented above, we need to determine whether a pattern $p_i$ visits regions of the attractor that are not visited by patterns already in $\mathrm{W}$.  If yes, then $p_i$ needs to be added to $\mathrm{W}$.  Such determination can be made using the {\em directed Hausdorff distance} \cite{rockafellar2009variational} between point sets: For any pair, A $\subset$ M and B $\subset$ M, of non-empty subsets of a metric space M the directed Hausdorff distance from A to B is defined as
\begin{equation}\label{eq:dHdAB}
d(\mathrm{A},\mathrm{B}) = \max_{a\in\mathrm{A}} \min_{b\in\mathrm{B}} \left\| a - b \right\|\,,
\end{equation}
where $\left\|\cdot\right\|$ denotes the distance between points in the metric space M.  Intuitively, $d(\mathrm{A},\mathrm{B})$ measures the largest distance from any point $a\in\mathrm{A}$ to the nearest point in B.  Therefore, if $d(\mathrm{A},\mathrm{B})$ is small, then every point in A is close to some point in B.  Note that, in general, $d(\mathrm{A},\mathrm{B}) \neq d(\mathrm{B},\mathrm{A})$.  

This is illustrated in Fig.~\ref{fig:dHdAB}, where M is a two-dimensional plane with the Euclidean distance between points.  We see that $d(\mathrm{B},\mathrm{A})$ is relatively small because the Euclidean distance from every point in B to the nearest point in A is small.  Conversely, $d(\mathrm{A},\mathrm{B})$ is relatively large because there are points in A whose distance to the nearest point in B is large.  Therefore, subset A is located in the vicinity of all parts of the subset B (or we could say 'covers all parts of B'), while the converse is not true.

The directed Hausdorff distance between a recurrent pattern $p_i(t)$ and a set of recurrent patterns $\mathrm{W} = \{w_j\}_{j=1}^{N_\mathrm{W}}$, which appears in Algorithm 1, is given by
\begin{equation} \label{eq:dHd}
d(p_i,\mathrm{W})=\max\limits_{t \in \left[0,T^{p_i}\right]}\min_{\substack{w_j \in\mathrm{W} \\ \tau\in \left[0,T^{w_j}\right]}} \left\|p_i(t)-w_j(\tau)\right\| ,
\end{equation}
where 
$\left\|\cdot\right\|$ denotes distance between points on the recurrent patterns.
Upon the completion of the above process of constructing W, all recurrent patterns contained in P will necessarily be within $\varepsilon$ distance of W, i.e.,~$d(\mathrm{P},\mathrm{W}) \leq\varepsilon$. Therefore, we say that W covers P with resolution $\varepsilon$.

Note that other notions of the 'distance' between recurrent patterns and/or segments of chaotic trajectories have been proposed in the literature.  For example, in the context of constructing recurrence networks from time series of chaotic dynamical systems, several measures based on distances of trajectory segments in the time-embedding space, or correlations between the embedded state vectors have been used~\cite{donner2010njp}, while in the context of coarse-grained representation of a turbulent flow by embedded periodic orbits, shape similarity of projected periodic orbits and segments of turbulent trajectories is determined using persistent homology methods and used as a measure of proximity~\cite{yalniz2021prl}.

\subsection{Distance in lower-dimensional projections}\label{sec:low-dim}
In order to distinguish close and distant trajectory segments using the directed Hausdorff distance, it is not necessary to work in the full phase space, which could be very high-dimensional (as, for example, in the case of the discretised Navier--Stokes flow).  A reliable distinction can be made in a low-dimensional projection of the space.  For example, one could take the first $n$ principal components of the attractor. 

For continuous-time systems and $n \geq 2$, $n$-dimensional projections of non-steady trajectories are curves, and therefore they are infinite-dimensional objects (i.e.,~they are points in an infinite-dimensional space of curves).  Generically, they do not coincide for two different trajectories (for coincidence, a special `non-generic' choice of projection is needed, e.g.,~with respect to some symmetries).  For discrete-time systems, we have to consider sufficiently long trajectories to guarantee this property. If projections of two trajectories never coincide, then the Hausdorff distance between projections is, at the same time, a distance in the space of trajectories.  This equivalence of spaces of trajectories and their projections is the essence of the famous Takens embedding theorem~\cite{Takens1981}.  The relationship between the dimension of the approximating attractor and the necessary length of the time series was considered by Eckmann and Ruelle~\cite{Eckmann92}.  Also, the Johnson-Lindenstrauss lemma~\cite{Johnson84}, which is widely used in computer science and signal processing, gives the boundaries for the preservation of the distances in random projections (up to a scalar scaling factor)~\cite{CandesTao06}.

Note that the values of distances differ in different projections.  Therefore, depending on which projection is used, both the number of recurrent structures needed to cover the attractor, and the specific structures chosen (for a given $\varepsilon$) will differ relative to the full-dimensional counterpart.  However, as we will demonstrate on the example of the spatiotemporally chaotic Kuramoto--Sivanshinsky flow, thanks to the metric equivalence mentioned earlier, a reliable reduced representation of the chaotic attractor by a minimal cover can be achieved using a projection with a relatively small number of dimensions.

\section{The Kuramoto--Sivashinsky Equation}\label{sec:kse}
To demonstrate the proposed method, we will use the set of recurrent patterns found by Cvitanovi\'{c}, Davidchack, and Siminos~\cite{cvitanovic2010state} for the Kuramoto--Sivashinsky (KS) equation
\begin{equation} \label{eq:ks}
u_t = -u_{xxxx}-u_{xx}-{\textstyle \frac{1}{2}} (u^2)_x\,,
\end{equation}
with periodic boundary conditions $u(x+L,t) = u(x,t)$ and system size $L=22$ -- a system with numerically stable spatiotemporally chaotic dynamics.  The subscripts $t$ and $x$ denote partial derivatives of $u(x,t)$ with respect to $t$ and $x$, respectively. The KS equation is a widely studied example of a relatively simple partial differential equation exhibiting spatiotemporal chaos. It is much simpler to analyse than, for example, the Navier--Stokes equations, yet exhibits complicated behaviour characteristic of turbulent dynamics~\cite{HolmesBook}. Derived independently by Kuramoto~\cite{kuramoto76} to study the angular-phase turbulence in the Belousov-Zhabotinskii reaction, Sivashinsky~\cite{sivashinsky77} to study the instability of laminar flame fronts, as well as several other authors~\cite{shkadov67,homsy74,nepomnyashchii74,laquey75}, this equation has become a cornerstone for many fundamental results in the study of spatiotemporal chaotic dynamical systems~\cite{kevrekidis1990back}.

As discussed in details in Ref.~\cite{cvitanovic2010state}, the KS equation (\ref{eq:ks}) is equivariant under translations (i.e.,~if $u(x,t)$ is a solution, then so is $\tau_s u(x,t) \equiv u(x+s,t)$) and reflections (if $u(x,t)$ is a solution, then so is $Ru(x,t) \equiv -u(-x,t)$).  Due to these symmetries, the set of recurrent patterns consist of equilibria ($u(x,t) = \hat{u}(x)$), relative equilibria or travelling waves ($u(x,t) = \hat{u}(x-ct)$ for any $t \in \mathbb{R}$, where $c$ is the speed of the travelling wave), and relative periodic orbits without reflection ($u(x,t+T) = \tau_s u(x,t) = u(x+s,t)$) and with reflection ($u(x,t+T) = R \tau_s u(x,t) = -u(-x-s,t)$), where $T$ is the orbit period. (The latter are also called {\em pre-periodic orbits}, since they are periodic with period $2T$: $u(x,t+2T) = R \tau_s u(x,t+T) = R\tau_s R\tau_s u(x,t) = R \tau_s \tau_{-s} R u(x,t) = u(x,t)$.)

Due to the periodic boundary conditions, the spectral method \cite{canuto1988book,trefethen2000book} is widely used to recast the partial differential equation (\ref{eq:ks}) in the form of an infinite system of ordinary differential equations. To this end, $u(x,t)$ is represented in terms of its Fourier expansion with respect to the spatial coordinate:
\begin{equation} \label{eq:ksfm}
  u(x,t)=\sum\limits_{k=-\infty}^{+\infty} a_k(t)e^{i q_k x},
\end{equation}
where $q_k = 2\pi k/L$ and $a_k(t) \in \mathbb{C}$ are the Fourier modes.   Since $u(x,t) \in \mathbb{R}$, $a_{-k}(t) = a_k^\ast(t)$, where the asterisk denotes complex conjugate.  Substituting (\ref{eq:ksfm}) into (\ref{eq:ks}) and using the orthogonality property of the Fourier modes, we obtain an infinite system of ordinary differential equations for $a_k(t)$:
\begin{equation} \label{eq:ksode}
\frac{d a_k}{dt} = (q_k^2 - q_k^4) a_k - i\frac{q_k}{2} \sum_{m = -\infty}^\infty a_m a_{k-m}\,,\quad k \in \mathbb{Z} \,.
\end{equation}
Because of the hyperviscous damping term $-u_{xxxx}$ in (\ref{eq:ks}), the magnitude of the Fourier modes quickly decays with $k$ for sufficiently large $k$.  Therefore, it is possible to obtain accurate solutions of the KS equation by truncating the Fourier series in (\ref{eq:ksfm}).  As shown in  \cite{cvitanovic2010state}, a sufficiently accurate solution of (\ref{eq:ks}) with $L = 22$ can be obtained using a truncated Fourier series with $|k| \leq 15$, resulting in a 30-dimensional dynamical system for real and imaginary parts of $a_k(t)$, $k = 1, 2, \ldots, 15$.  (We see from ($\ref{eq:ksode}$) that $d a_0 /dt = 0$, so that $a_0(t) = const$.  Due to the so-called Galilean invariance of the KS equations, we can set $a_0(t) = 0$ for all $t$ without loss of generality \cite{cvitanovic2010state}).  This system can be efficiently solved using the exponential time-differencing fourth-order Runge–-Kutta method (ETDRK4) \cite{cox2002jcop,kassam2005sjsc}, with accurate solutions obtained using the integration stepsize $h = 0.25$.

Excluding the trivial equilibrium $u(x,t) = 0$, the set of detected recurrent patterns for this system, $\mathrm{P}=\left\{p_i\right\}$, consists of three equilibria ($i = 1,2,3$), two travelling waves ($i = 4,5$), and over $60\,000$ relative periodic orbits and pre-periodic orbits with periods $T < 200$,  \cite{cvitanovic2010state}, ($i > 5$, orderer by increasing orbit period $T^{p_i}$).  This will be the set from which we will select the minimal cover set W using the algorithm described in Section \ref{sec:algorithm}. 

\subsection{Minimal cover for the KS attractor} \label{sec:kscover}
A unique representation of the KS dynamics on a periodic domain (\ref{eq:ks}) requires the use of symmetry-reduced coordinates, so that all solutions related by the above described symmetries coincide.  In general, the construction of a symmetry-reduced representation of a given dynamical system is not straightforward.  For (\ref{eq:ks}), one such representation was proposed in Ref.~\cite{budanur15prl}.  However, the issue with this and similar symmetry-reduction approaches is that points that are 'close' in the original phase space of the dynamical system can become 'distant' in the symmetry-reduced coordinates.  In principle, one could attempt to define an appropriate metric in the symmetry-reduced space, but this again does not appear to be straightforward.  

On the other hand, the use of the Hausdorff distances between segments of trajectories instead of the distances between points in phase space allows us to choose any convenient symmetry-invariant coordinates for the construction of the minimal cover.  Even though such choice of coordinates may identify points in phase space that are not related by the symmetries, the trajectories starting from such points will be different, with non-zero Hausdorff distance between them.  In the current work, we will use the magnitudes of the complex Fourier modes, $|a_k|$, as the symmetry-invariant coordinates.  Even though in this representation we do not distinguish points in phase space with the same Fourier mode magnitudes and different phases, the fact that we are looking at segments of trajectories, rather than points, allows us to faithfully distinguish segments starting from different initial points in the full phase space.  In fact, as we demonstrate below, we can even take a subset of $|a_k|$ and still make such a distinction.  In the language of Section~\ref{sec:low-dim}, we will use projections of KS solutions on a lower-dimensional space described by a subset of Fourier magnitudes $|a_k|$ with selected values of $k$.

In order to decide which $k$ values to use, we determine the size of the chaotic attractor along each of the Fourier modes by measuring the maximum magnitudes $m_k = \max_t | a_k(t) |$ along a long trajectory in the attractor, $t \in [0, 10^7]$.  The magnitudes of the first seven Fourier modes are $m_{1, \ldots, 7}~=~\{0.5714, 0.9807, 1.2386, 0.5786, 0.3264, 0.2238,$ $0.0763\}$, with the values of $m_k$ for $k > 7$ exponentially decreasing approximately as $40\mathrm{e}^{-0.87k}$.  For the $n$-dimensional projection of the phase space we use the $n$ Fourier modes with the largest values of $m_k$.  Thus, the 2-dimensional projection will include modes 2 and 3, the 3-dimensional projection will include modes 2, 3 and 4, and from then onwards for the $n$-dimensional projection the first $n$ modes are used.  We will denote W$_n$ the minimal cover set constructed in the $n$-dimensional symmetry-invariant projection.

In what follows, the Euclidean distance between phase space points $x$ and $y$ in the $n$-dimensional projection is denoted as $\left\| x - y \right\|_n$ and the directed Hausdorff distance in the same projection is denoted $d_n(\cdot,\cdot)$. Recall that in the construction of the minimal cover, see Algorithm~1, the threshold parameter $\varepsilon$ determines the resolution of the cover.  In the $n$-dimensional projection, we will use the value of $\varepsilon_n$ equal to 1\% of the length of the main diagonal of the hypercuboid with sides $m_{k_j}$, $j = 1, \ldots, n$, where $k_j = \{3, 2, 4, 1, 5, 6, 7, \ldots, 15\}$; that is, $\varepsilon_n = 0.01\sqrt{\sum_{j=1}^n m_{k_j}^2}$, resulting in the values $\varepsilon_{2,\ldots,7}~\approx~\{.01579, .01682, .01776, .01807, .01820, .01822\}$, and $\varepsilon_{15} \approx .01822$.

\begin{figure}[tb]
\begin{center}
\includegraphics[width=0.9\textwidth]{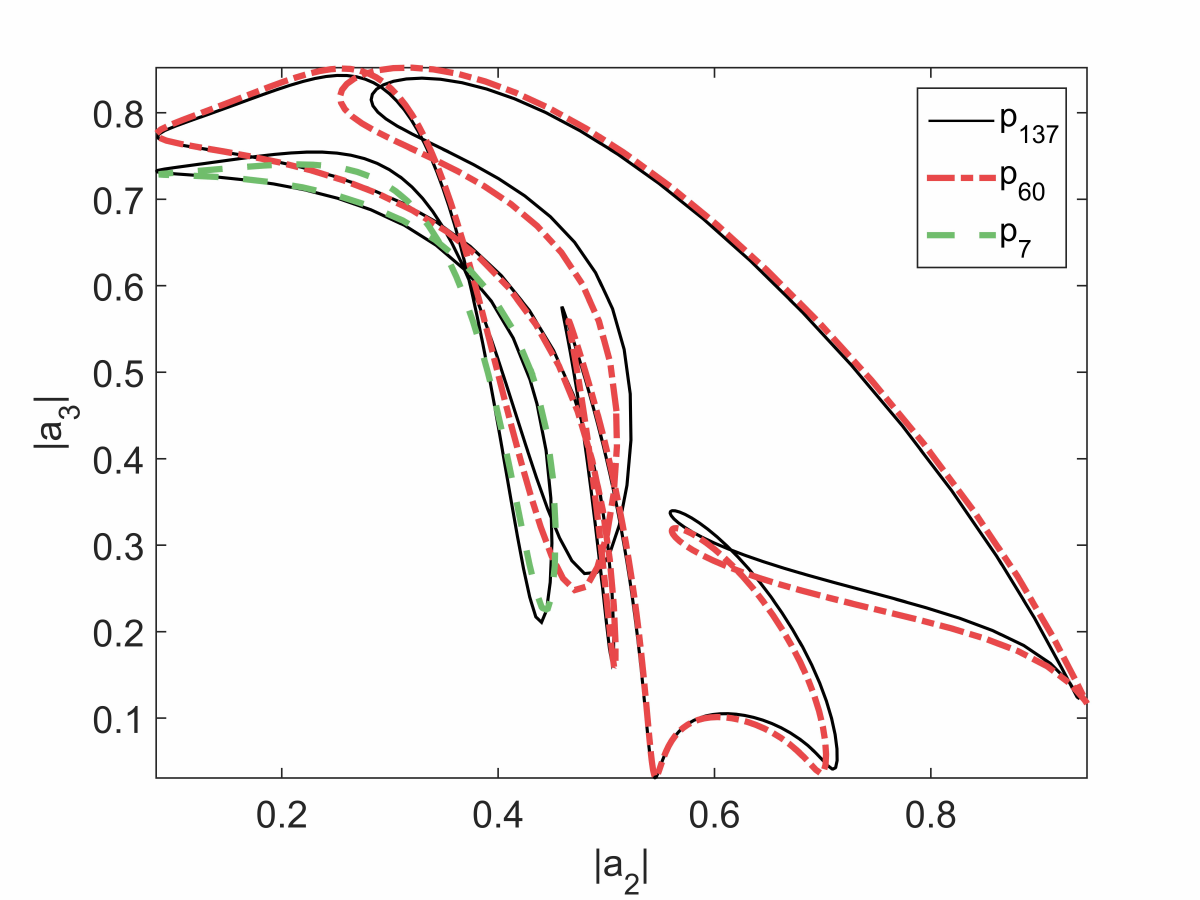}
\end{center}
\caption{\label{figP66} Shadowing of the recurrent pattern $p_{137}\notin \mathrm{W}_4$ by $p_{7}, p_{60} \in \mathrm{W}_4$.}
\end{figure}

The application of the above outlined construction procedure to the set P in the $n$-dimensional symmetry-invariant projections resulted in the minimal cover sets W$_n$ containing 91, 513, 747, 809, 825, and 830 recurrent patterns for $n = 2, \ldots, 7$, respectively. We can see that the minimal cover sets with 1\% resolution contain only a small fraction of the over 60\,000 available recurrent patterns from the set P. In other words, all recurrent patterns in P can be approximated (or `shadowed') with $\varepsilon_n$ resolution by those within the small subset, W$_n$. An example of such shadowing is shown in Fig.~\ref{figP66}, where a relative periodic orbit $p_{137}$ not in W$_4$ with period 76.64 is shadowed by two relative periodic orbits in W$_4$ with periods 62.95 and 14.33.

In order to ascertain how well each W$_n$, $n = 2, \ldots, 7$, covers the full set of recurrent patterns P in the 15-dimensional space of all Fourier mode magnitudes, we calculated the percentage of recurrent patterns in the sets $\mathrm{P}\setminus\mathrm{W}_n$ that can be fully covered by W$_n$ with directed Hausdorff distances calculated in the 15-dimensional symmetry-invariant space.  In other words, what fraction of patterns $p_i \in \mathrm{P}\setminus\mathrm{W}_n$ satisfy: $d_{15}(p_i,\mathrm{W}_n)\leq\varepsilon_{15}$?  We find that for $n=2,\ldots,7$ the fraction of patterns satisfying this condition are: $0.008, 0.785, 0.995, 0.998, 0.999, 1$, respectively. Upon further investigation, we also found that, for $n \geq 4$, this fraction increases to $1$ if we consider patterns that have at least 99\% of their trajectory covered by W$_n$ in the 15-dimensional symmetry-invariant space.  From this we can conclude that the smallest dimension of the projection that can be used to construct a reliable minimal cover is $n = 4$.  Therefore, in what follows, we will use W$_4$ for the analysis of shadowing of the chaotic attractor by the minimal cover set of recurrent patterns.


\subsection{Shadowing of the chaotic attractor}\label{sec:shadowing} 
With the help of the constructed minimal cover set W$_4$, we can now begin to put Poincar\'{e}'s conjecture into practice by representing a chaotic trajectory as a sequence of segments shadowed by recurrent patterns in W$_4$.

Note that the problem of shadowing of a chaotic attractor is quite subtle.  If the attractor is connected and transitive, then this problem is equivalent to shadowing of long chaotic trajectories.  If it is not, then shadowing of a trajectory will approximate only a transitive connected part of the attractor to which this trajectory belongs. On the other hand, $\varepsilon$-trajectories might shadow a wider set than the attractor even when $\varepsilon \to 0$. The asymptotic theory of attractors of  $\varepsilon$-motions when $\varepsilon \to 0$ was developed in 1980 in the thesis of A.~N.~Gorban (see partial English translation in~\cite{gorban1980phd} or~\cite{gorban1997arxiv}).  

If an attractor is hyperbolic, then shadowing of a chaotic trajectory can be achieved over an arbitrarily long time interval~\cite{katok95book}. In nonhyperbolic systems, shadowing of a trajectory can extend only over a finite time interval~\cite{grebogi2002}. In our current work, when discussing shadowing of a chaotic attractor, we assume that all parts of the attractor are shadowed by some recurrent pattern in the available set P. If this is not the case, then we are talking about shadowing only that part of the attractor that has this property.

Recall that our goal is to construct a cover of the chaotic set by $\mathrm{W}_n$ that is as good as the cover by the full available set P.  Now that we have the necessary definitions, we can state more precisely what we mean by this goal.  Note that P covers with the $\varepsilon_n$ resolution all regions of phase space $\mathcal{M}$ which are within the $\varepsilon_n$-ball of P: 
\begin{equation}\label{eq:ball}
B_{\varepsilon_n}(\mathrm{P}) = \{x \in \mathcal{M} : d_n(x,\mathrm{P}) < \varepsilon_n \},
\end{equation} 
where 
\begin{equation}\label{eq:dn}
d_n(x,\mathrm{P}) = \min_{\substack{p_i \in\mathrm{P} \\ t \in \left[0,T^{p_i}\right]}} \left\|x - p_i(t)\right\|_n\,.
\end{equation}  
Intuitively, $B_{\varepsilon_n}(\mathrm{P})$ consists of all point in phase space $\mathcal{M}$ whose distance to the nearest point in P is smaller than $\varepsilon_n$.  Also, by construction of the minimal cover W$_n$, P $\subseteq B_{\varepsilon_n}(\mathrm{W}_n)$ and so $B_{\varepsilon_n}(\mathrm{P}) \subseteq B_{2\varepsilon_n}(\mathrm{W}_n)$, i.e., the minimal cover $\mathrm{W}_n$ with resolution $2\varepsilon_n$ provides the same cover as P with resolution $\varepsilon_n$.  Therefore, when investigating the shadowing of the chaotic attractor by the minimal cover $\mathrm{W}_n$, it is appropriate to use the $2\varepsilon_n$ resolution.

There are many possible ways to associate a segment of a chaotic trajectory with a particular recurrent pattern within W$_4$.  For example, we can use the `nearest' representation, where each point on the chaotic trajectory is associated with the nearest pattern in W$_4$.  However, because we are working with the lower dimensional projections of the dynamics, such representation may not always find the nearest pattern in the full space.  Another approach would be to find all recurrent patterns which are within $2\varepsilon_4$ of a chaotic trajectory segment and associate the segment with one such structure chosen at random.  Such a `random' representation would be more appropriate for systems with noise, for example when analysing experimental time series data.

In the present work, we adopt a `greedy' representation, where from among all the recurrent patterns in W$_4$ that are within $2\varepsilon_4$ of a chaotic trajectory segment, we associate the segment with the pattern which stays within $2\varepsilon_4$ for the longest period of time.  This approach is consistent with our earlier observation that segments of trajectories that remain close to one another in a low-dimensional projection for a significant time interval, are likely to be close, or `dynamically linked', in the full phase space.

More specifically, given a chaotic trajectory $\mathbf{a}(t) = \{a_1(t), a_2(t), \ldots, a_{15}(t)\}$, $t \ge t_0$ with the starting point $\mathbf{a}(t_0)$, we find the pattern $w_{s_0}\in$ W$_4$ that stays within $2\varepsilon_4$ of the trajectory for the longest period of time, say until $t = t_1$.  Then we associate the chaotic trajectory segment $\phi_0 = \{\mathbf{a}(t), t \in [t_0, t_1)\}$ with the recurrent pattern $w_{s_0}$.  Repeating this process from the point $\mathbf{a}(t_1)$, we associate the segment $\phi_1 = \{\mathbf{a}(t), t \in [t_1, t_2)\}$ with the recurrent pattern $w_{s_1}$.  And so on.  As a result, the chaotic trajectory is subdivided into finite time segments $\{\phi_0, \phi_1, \ldots\}$ labelled with indices $\{s_0, s_1, \ldots \}$ such that segment $\phi_i$ is shadowed by $w_{s_i} \in \mathrm{W}_4$.

\begin{figure}[tb]
\begin{center}
\includegraphics[width=0.9\textwidth]{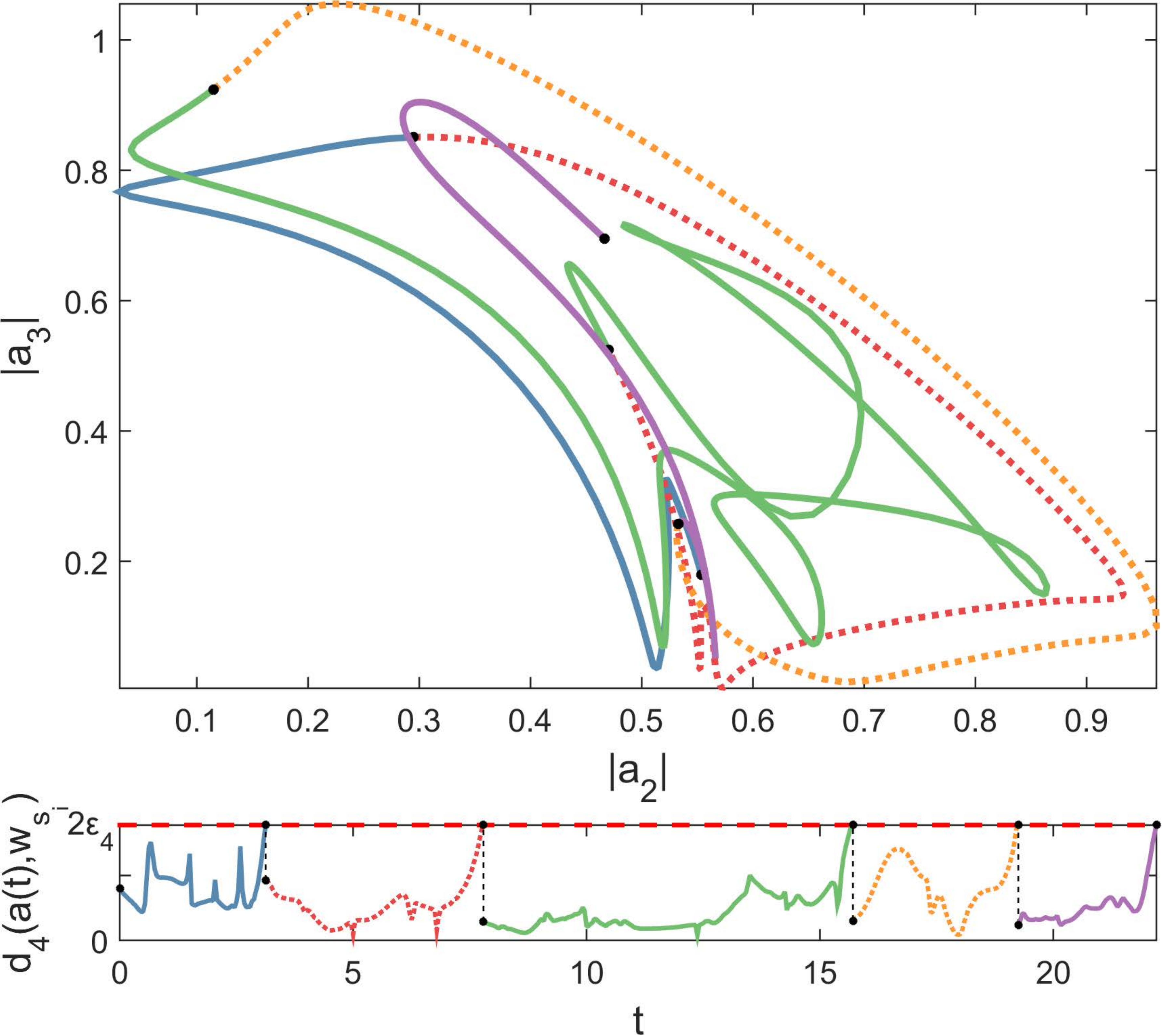}
\end{center}
\caption{\label{figShadTraj} {\em Top}: Encoding of a short segment of a typical chaotic trajectory in the attractor by patterns in W$_4$, shown in a two-dimensional projection. {\em Bottom}: Distance between trajectory and chosen pattern in W$_4$ at time $t$.  The chosen patterns for each respective segment are $w_{8}, w_{237}, w_{242}, w_{70},$ and $w_{319}$; segment transitions are indicated by black dots and a change in the line colour and style.}
\end{figure}

In Fig.~\ref{figShadTraj} we show an example of the encoding of a typical chaotic trajectory segment $\mathbf{a}(t), t\in[0,22.21]$, by W$_4$ using a `greedy' shadowing algorithm.  In the upper part of the Figure, $\mathbf{a}(t)$ is depicted in projection onto $|a_2|$ and $|a_3|$ with the different segments marked by different line styles and colours.  In the lower part of the Figure is the distance between the trajectory segment and the assigned recurrent pattern in W$_4$. We see that, as expected, when $d_4(\mathbf{a}(t),w_{s_i})=2\varepsilon_4$ the algorithm finds the next pattern in W$_4$ that stays within $2\varepsilon_4$ for the longest period of time.

\section{Markov model approximation}\label{sec:markov}

The idea of Markov models for complex dynamical systems was introduced in the late 1960's and early 1970's by Sinai~\cite{sinai2020markov}, Bowen~\cite{bowen1970markov}, Adler and Weiss~\cite{adler1970similarity}, and others. The goal of this approach was to find new dynamical invariants for chaotic attractors and to prove non-equivalence of different chaotic dynamical systems.  The notion of Markov partition was introduced primarily for theoretical analysis of hyperbolic dynamical systems.  Nevertheless, we should state that the Markov model is not only an excellent instrument for theoretical analysis, but also an important applied tool.  Potentially, there are many approaches for Markov chain approximation of dynamical systems.  In this work, we consider a Markov model consisting of motion along segments of trajectories with small random jumps between them.  To construct such a model, we need to answer two questions: 1) How to create the set of segments of trajectories that correspond to different states of the Markov model? 2) How to calculate the probabilities of jumps, or transitions, between them.  With the introduction of the minimal cover we have addressed the first question, and now we turn our attention to the second question.

Representation of the dynamics on the attractor by a sequence of transitions between cover elements opens up the possibility of approximating the dynamics by a Markov process. In order to do so, we followed a long ($t = 10^{10}$) orbit on the chaotic attractor, segmenting it into transitions between cover elements in W$_4$ and waiting (also called residence or holding) times at each element.  This information is used to construct the finite-state continuous-time Markov Chain (CTMC) model, as described below.


\subsection{Construction}

In order to construct the transition probability matrix for the CTMC, one must first construct the transition \emph{rate} matrix (also known as the \emph{generator}) of the Markov chain. The transition rate matrix \textbf{Q} takes the form:
\begin{equation} \label{eq:transratematrix}
\mathbf{Q} = 
\begin{bmatrix}
    -q_{1}       & q_{12} & q_{13} & \dots & q_{1n} \\
    q_{21}       & -q_{2} & q_{23} & \dots & q_{2n} \\
    \vdots & \vdots & \vdots & \ddots & \vdots \\
    q_{n1}       & q_{n2} & q_{n3} & \dots & -q_{n}
\end{bmatrix},
\end{equation}
where $n$ is the number of states.  We have investigated two methods for estimation of the generator from the transition data, as discussed in Ref.~\cite{metzner2007generator}.  The first, based on the \emph{Quadratic optimisation} method, was first introduced by Crommelin and Vanden-Eijnden~\cite{crommelin2006fitting}. This method involves calculating an estimate for the generator through the solution of a quadratic minimisation problem. One of the drawbacks of this method, however, is that, if the number of states $n$ is too large to store the tensor $H\in\mathbb{R}^{n\times n \times n \times n}$, then the problem must first be reformulated \cite{crommelin2009data}.

Alternatively, there is the \emph{Expectation maximisation} method introduced by Holmes and Rubin \cite{holmes2002expectation} which involves using data to find an estimate of the generator such that the given continuous log-likelihood function is maximised. While Holmes and Rubin first introduced this method as a tool for estimating a generator for continuous-time observations, Bladt and S{\o}rensen \cite{bladt2005statistical} later extended the method to allow for the construction of an approximate generator using data observed at discrete time intervals, as in our case.

Metzner et al.~\cite{metzner2007generator} analysis concluded that, in most cases, the difference between the quadratic programming and expectation maximisation methods is negligible, with a slight edge given to the quadratic programming method. Therefore, because of its relative simplicity, we have used the expectation maximisation method for the construction of the approximate generator.  The construction of the transition rate matrix \textbf{Q} uses the maximum likelihood estimates of $q_{ij}$ defined as $\hat{q}_{ij} = \tfrac{N_{ij}}{R_{i}}$, where $N_{ij}$ is the number of transitions from state $i$ to state $j$ and $R_{i}$ is the total time spent in state $i$ \cite{bladt2005statistical}.  The diagonal elements $q_{i}$ are then defined as the sum of all non-diagonal elements of each row, i.e.,~$q_i = \sum_{i\neq j}  q_{ij}$, so that each row of \textbf{Q} in (\ref{eq:transratematrix}) sums to zero.

Having calculated \textbf{Q}, the transition probability matrix \textbf{A} can be computed using the matrix exponential:
\begin{equation}
\mathbf{A}(t) = e^{\mathbf{Q} t},
\end{equation}
where \textbf{A}$_{ij}(t)$ is the probability of observing a transition from state $i$ to state $j$ within time interval $t$ \cite{bladt2005statistical}. Note that when we write simply \textbf{A} in what follows, we refer to the transition probability matrix after unit time, $\mathbf{A}(1)$.

In order to construct the transition probability matrix $\mathbf{A}$ for our system, we first performed shadowing on a trajectory of duration approximately $10^{10}$ using W$_4$. By applying the above procedure, the result is the $747 \times 747$ matrix $\mathbf{A}$.

\subsection{Properties}

\begin{figure}[t]
\includegraphics[width=0.9\textwidth]{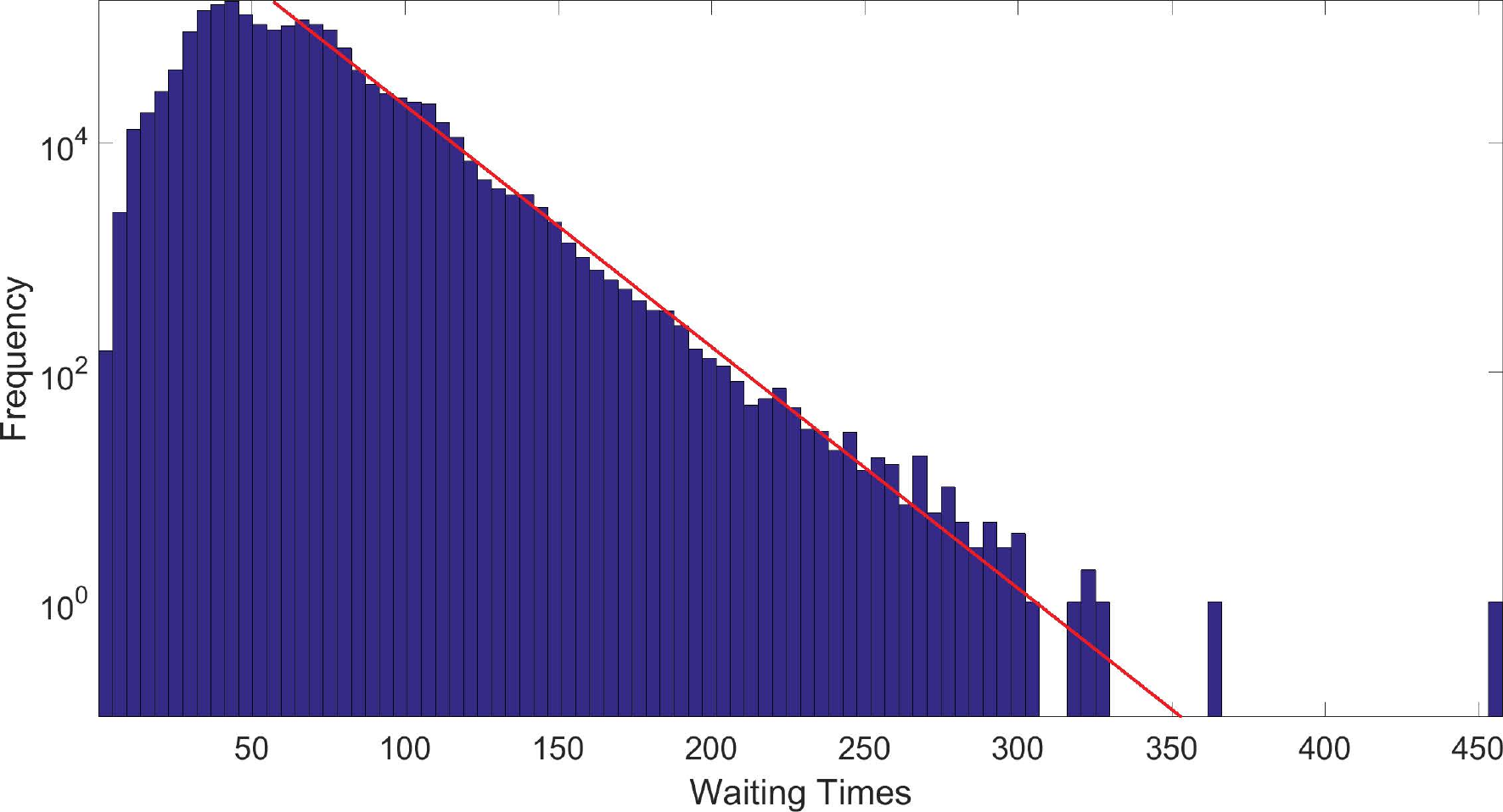}
\caption{\label{figWaitingTime} Distribution of waiting times on a chaotic trajectory of duration $10^{10}$ shadowed by recurrent patterns in the minimal cover W$_4$. The straight red line is the least-squares fit in the interval $[60, 300]$.}
\end{figure}

In order to see how well dynamics on the KS attractor can be modelled by a CTMC model, we can check to what extent the model exhibits the memoryless property~\cite{norris1998markov}, which manifests itself in the exponential distribution of the waiting times~\cite{nicolis1997markov,gillespie1991markov}.  As can be seen in Fig.~\ref{figWaitingTime}, we do observe an approximately exponential distribution for waiting times larger than about 70. The straight red line is obtained by the least-square fit to the data in the interval $[60, 300]$ and is approximately equal to $\mathrm{e}^{14.5-0.048 t_w}$, where $t_w$ is the waiting time.  For smaller waiting times the distribution is not exponential because the system is deterministic, thus strongly correlated on shorter time scales.  However, due to sensitive dependence of chaotic dynamics on initial conditions, the memory of the initial condition is lost after a sufficiently long time interval. This feature of chaotic systems is discussed in detail in many different contexts, including research into distinguishing chaotic processes from stochastic processes/noise~\cite{kantz2004fast,cencini2000chaos}, and also in the context of the applicability of Markov models to chaotic systems~\cite{nicolis1990chaotic,nicolis1997markov}.  The straight-line fit in Fig.~\ref{figWaitingTime} also indicates a heavier than exponential tail of the distribution for waiting times larger than about 250, but the statistical significance of this observation is not sufficient to draw a definite conclusion.

\subsubsection{Stationary Distribution} \label{sec:statdist}

\begin{figure}[ht]
\centering
\includegraphics[width=\textwidth]{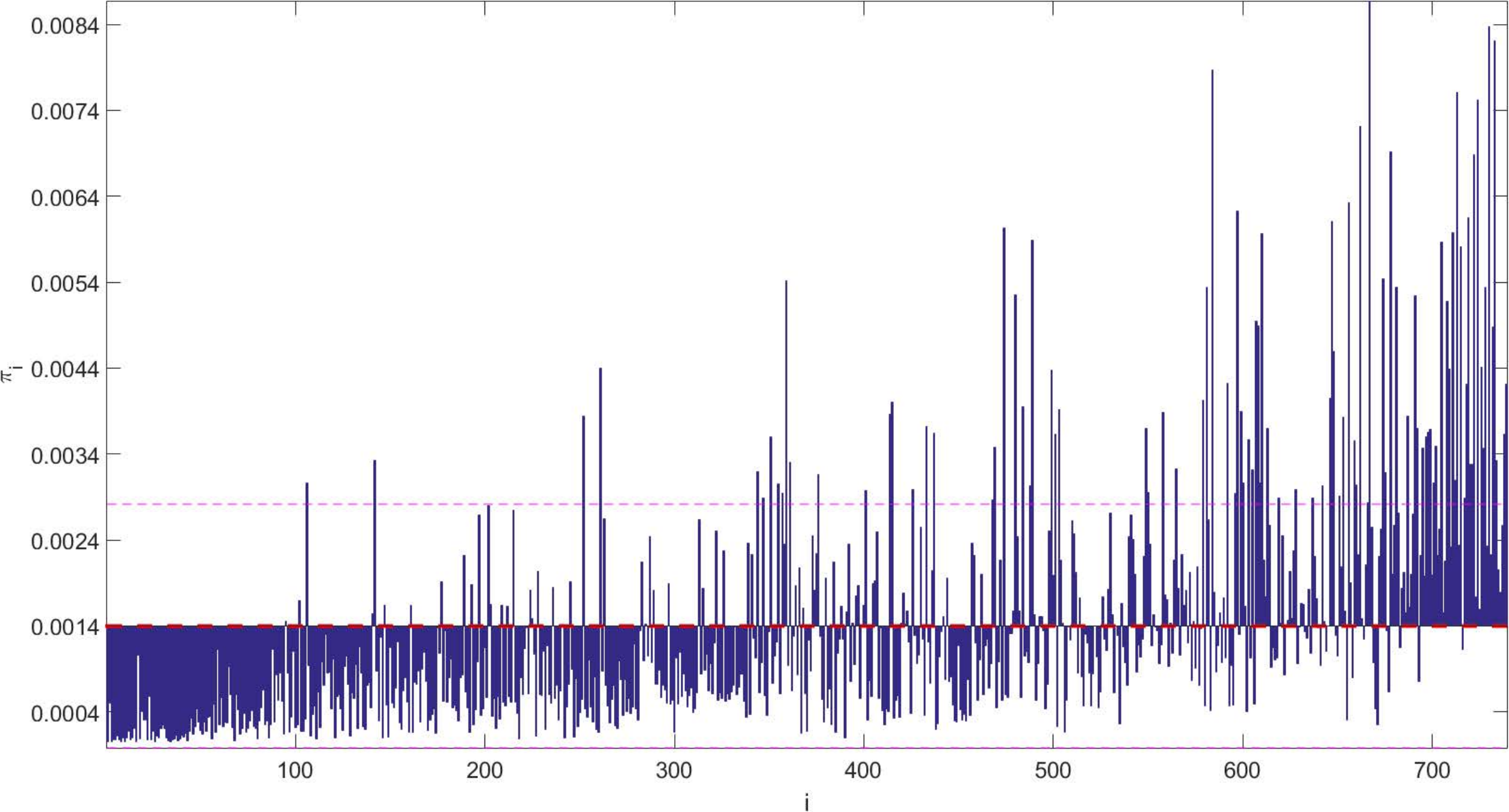}
\caption{$\pi_i$ of each cover element, with the middle red dashed line corresponding to the mean $\mu_{\pi}$ and the outer magenta dashed lines corresponding to one standard deviation above and below the mean ($\mu_{\pi}\pm \sigma_{\pi}$).}
\label{fig:kse_pi_stdev} 
\end{figure}

The stationary distribution $\pi$ of a Markov chain is the left eigenvector of $\mathbf{A}$ corresponding to eigenvalue 1, $\pi \mathbf{A} = \pi$, satisfying the normalisation condition $\sum_i \pi_i = 1$.  In order for the stationary distribution to be unique, the Markov chain must be \emph{ergodic} -- meaning that it is possible to reach any state from any other state within a finite time~\cite{levin2009markov}. When analysing the waiting times, we found that cover elements 1--8 in W$_4$ were not used at all during the shadowing of the long chaotic trajectory.  In the case of the equilibrium (patterns 1, 2, and 3) and travelling wave solutions (patterns 4 and 5), this is due to the fact that they lie on the outskirts of the attractor, so it could be expected that they are visited very rarely.  In the case of the recurrent patterns 6, 7, and 8, which are short relative periodic orbits, all points on these orbits are completely covered by points from other, longer orbits; as such, the shadowing algorithm chooses the longer orbits in place of these shorter ones, as the chaotic trajectory appears to follow these longer orbits for more time.  By removing states 1--8 from the Markov model we reduce $\mathbf{A}$ to a $739 \times 739$ matrix, representing an \emph{ergodic} Markov chain.

The stationary distribution $\pi$ gives the probability that the chaotic trajectory is shadowed by a particular state at any given time, and so by analysing these probabilities we can determine the relative importance of different recurrent patterns to the dynamics on the chaotic attractor.  Fig.~\ref{fig:kse_pi_stdev} shows the probabilities $\pi_i$ for each cover element $i$ as deviations from the mean $\mu_{\pi}$, with one standard deviation above and below the mean indicated by the magenta dashed lines.  From this Figure we see that the shorter recurrent patterns play a generally less important role in the dynamics.  One of possible explanations is that, when new longer-period patterns are added to W$_n$ in the process of its construction, these patterns also provide cover for the regions of the attractor covered by the earlier added shorter patterns.  We will explore this issue in the next Section.

%

\section{Redundancies} \label{sec:redun}

\begin{figure}[h]
\includegraphics[width=0.85\textwidth, keepaspectratio]{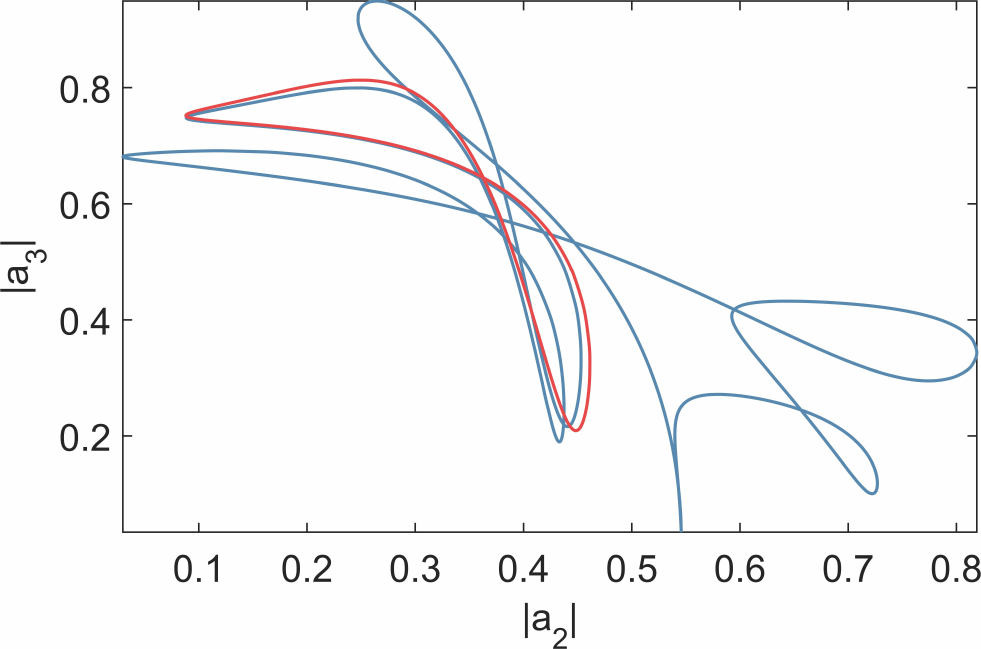}
\caption{\label{fig:redundancyExample} Example of redundancy in W$_4$: $w_{8}$ (red) is completely within $\varepsilon_4$ distance of $w_{53}$ (blue).}
\end{figure} \hfill

Due to the ordered nature of the construction of the minimal cover proposed in Section~\ref{sec:kse} (ordered from shorter to longer recurrent patterns), it is possible that a recurrent pattern that is included into the set early (say, $w_i$) could be completely covered by the one added later (say, $w_j$, $j>i$).  If $w_j$ completely covers $w_i$, and explores some additional part of the attractor, then $w_i$ would (by construction) not cover $w_j$. In other words, it is possible that $d(w_i,w_j)\leq\varepsilon$ while $d(w_j,w_i)>\varepsilon$,
in which case the covering provided by W$\setminus w_i$ is almost the same as that provided by W.  An example of this can be seen in Fig.~\ref{fig:redundancyExample}, where (for W$_4$) $w_{8}$ is completely covered by $w_{53}$, which also explores additional parts of the chaotic attractor.  Since $d_4(w_{8},w_{53})\leq\varepsilon_4$, and it is clearly observable that $d_4(w_{53},w_{8})>\varepsilon_4$, $w_{8}$ can be considered redundant in that almost no covering would be lost by removing it from the minimal cover set.

\subsection{Second Pass}

With this in mind, we add another stage to the construction of the minimal cover, by which such redundancies can be removed from the set.  Having obtained the minimal cover W using Algorithm~1, we introduce a {\em second pass} through the recurrent patterns in W, as presented in Algorithm~2.

\begin{algorithm}\label{alg:cover2}
\caption{Second pass to remove redundancies in the cover set W}
\begin{algorithmic}
\Require $\mathrm{W}=\left\{w_i\right\}_{i=1}^{N_\mathrm{W}}$; ~~$\varepsilon > 0$
\State $\overline{\mathrm{W}} \gets $ W
\For{$i = 1, \ldots, N_\mathrm{W}$}
  \If{$d(w_i,\overline{\mathrm{W}}\setminus w_i)\leq\varepsilon$}
    \State $\overline{\mathrm{W}} \gets \overline{\mathrm{W}} \setminus w_i$
  \EndIf
\EndFor

\hspace*{-1em}\Output $\overline{\mathrm{W}}$
\end{algorithmic}
\end{algorithm}

Intuitively, this second pass goes through all recurrent patterns in W and removes those that are covered by the subsequent patterns in W. One drawback of this method is that by removing recurrent patterns satisfying $d(w_i,\mathrm{W}\setminus w_i)\leq\varepsilon$, it is possible that we may lose some of the covering provided by the $\varepsilon$-neighbourhood of $w_i$. Indeed, the cover after the second pass, $\overline{\mathrm{W}}$, satisfies: $B_{\varepsilon}(\overline{\mathrm{W}}) \supseteq \mathrm{W}$.  This means that, if we wish to provide the same covering as P, we need to take $B_{2\varepsilon}(\overline{\mathrm{W}})\supseteq B_{\varepsilon}(\mathrm{W}) \supseteq \mathrm{P}$; further, if we wish to provide the same covering of the attractor as $B_{\varepsilon}(\mathrm{P})$ (as discussed previously), then we must take a $3\varepsilon$-ball around $\overline{\mathrm{W}}$: $B_{3\varepsilon}(\overline{\mathrm{W}})\supseteq B_{2\varepsilon}(\mathrm{W}) \supseteq B_{\varepsilon}(\mathrm{P})$. As such, the resolution of the cover $\overline{\mathrm{W}}$ is further somewhat reduced by applying the second pass.

The result of applying the second pass to our previously constructed covers W$_{2,\ldots,7}$ using the same $\varepsilon_{2,\ldots,7}$ as previously used are new covers $\overline{\mathrm{W}}_{2,\ldots,7}$ containing 29, 201, 306, 333, 339, and 342 coherent structures respectively, an over 50\% reduction in the size of each cover set.  When checking to what extent the minimal set $\overline{\mathrm{W}}_n$ covers the set P$\setminus \overline{\mathrm{W}}_n$ in the 15-dimensional symmetry-invariant subspace, we determined that, for $n \geq 4$, all recurrent patterns are covered for at least 98\% of their period, which is a reduction from the 99\% we observed for the minimal cover sets W$_{4,\ldots,7}$.  As such, we see that the minimal cover after the second pass still provides an adequate cover with over 50\% fewer recurrent patterns.

It is interesting to note that the second pass removes a large fraction of shorter recurrent patterns because they are fully covered by the longer ones.  This goes against the intuition developed from the periodic orbit theory, where shorter periodic orbits are forming the fundamental structure of the 'skeleton of chaos', with longer orbits providing small refinements of this structure~\cite{cvitanovic1991periodic}.  At the same time, this is consistent with our observations in Section~\ref{sec:statdist} (see Fig.~\ref{fig:kse_pi_stdev}) that longer recurrent patterns are visited more often by a long chaotic trajectory compared to the shorter ones.  As such, relatively few long recurrent patterns covering all parts of a chaotic attractor appear to be quite efficient at representing the properties of the chaotic attractor.

\section{Summary}\label{sec:conclude} 
We have presented a general method for constructing a minimal cover of a high-dimensional attractor from the known recurrent patterns embedded in the attractor. The key element of this construction is the use of the directed Hausdorff distance between segments of trajectories in projection on a suitably chosen set of coordinates of the dynamical system, which allows us to determine the proximity of recurrent patterns to one another or to a chaotic trajectory. We have given an example of such a construction for a spatiotemporally chaotic Kuramoto--Sivashinsky chaotic attractor and explored the effect that the dimension of the projection used during the construction has on the covering of the attractor.  We show that a reliable minimal cover can be constructed in a low-dimensional projection of the attractor, thus substantially reducing the complexity and computational cost of such a construction compared to the full phase space representation. Here we should mention a deep analogy between selection of the minimal cover set and selection of the prototypes in the condensed nearest neighbour algorithm (or the Hart algorithm~\cite{hart1968}). It was designed to reduce the datasets for $k$ nearest neighbour ($k$NN) classification by selecting the set of prototypes U such that 1NN with U can classify the examples almost as accurately as 1NN does with the whole data set.  We have also demonstrated how a typical trajectory in the attractor can be encoded as a sequence of cover elements which shadow this trajectory.  This opens up the possibility of describing the dynamics on the attractor as a Markov-type model with a relatively small number of states.  Such a model allows to investigate the structure of high-dimensional chaotic dynamics, which will be the subject of future research.

Finally, we would like to note that the minimal cover construction can be based not just on recurrent patterns, but on any convenient set of segments of trajectories from a chaotic attractor. As such, it represents a generic tool for constructing a coarse-grained representation of a high-dimensional chaotic attractor, where different elements of the minimal cover set represent dynamics of different parts of the attractor with minimal redundancy.  For many applications, such a coarse-grained representation may be as informative as that based on recurrent patterns, which may not be available, or very costly to locate. To what extent the quality of the coarse-grained representation of a high-dimensional chaotic attractor depends on the use of recurrent patterns as opposed to generic trajectory segments will be the subject of future research.


\end{document}